\documentclass[aps,prl,twocolumn,groupedaddress]{revtex4}

\usepackage{graphicx}
\usepackage{dcolumn}
\usepackage{color}
\usepackage{amssymb}

\begin{document}

\title{Thermodynamical Signatures of an Excitonic Insulator}

\author{Benno Bucher}
\affiliation{HSR Hochschule f\"{u}r Technik, 8640 Rapperswil, Switzerland}

\author{Tuson Park, J. D. Thompson}
\affiliation{Los Alamos National Laboratory, Los Alamos, NM 87545, USA}

\author{Peter Wachter}
\affiliation{ Lab. f\"{u}r Festk\"{o}rperphysik, ETH Z\"{u}rich,  8093 Z\"{u}rich, Switzerland}

\date{\today}
\begin{abstract}
In the 1960s speculations arose if a ground state exists in solid state materials with an electron
and a hole bound to a pair with their spins added to integer values, i.e. excitons. Here we show
that electrons and holes in TmSe$_{0.45}$Te$_{0.55}$ do form excitons as the thermodynamical ground
state. The formation of a large number of excitons in an indirect gap semiconductor requires
momentum conservation by means of phonons and, hence, implies a significant change of the heat
capacity of the lattice, as found experimentally. The thermodynamically derived phase diagram
sustains a bosonic ground state in condensed matter.
\end{abstract}


\maketitle


The band structure of metals, semiconductors and insulators is the consequence of the fermionic character of the electrons. However, if an
electron is excited out from the valence band, a positively charged hole stays behind, attracting the negatively charged electron due to the
Coulomb attraction. A bound electron-hole pair is called an exciton; such a quasiparticle is a boson since the individual spins add to spin zero
or one (see Fig. 1). Mott \cite{mott} suggested that a semimetal (SM) is unstable against the formation of such a bound electron- hole pair
because the small density of free electrons does not screen the Coulomb attraction. Knox \cite{knox} realized that the band structure of a
semiconductor (SC) might also be unstable against the spontaneous formation of excitons if the binding energy of the exciton is larger than or
comparable to the energy gap. The notion of the excitonic insulator (EI) has been coined.

The  elaboration of the theoretical concept of excitonic insulators provided insights into the possible mechanism of how a new ground state
could form \cite{rice}. Weakly bound excitons should reveal a simultaneous formation and phase coherence,  as within BCS theory; whereas,
strongly coupled excitons form at high temperatures and are subject to a Bose condensation (BC) at a lower temperature. The crossover from weak
to strong coupling became essential for understanding the high temperature superconductors. The Bose statistics of the excitons would initiate
new features and promise new effects and applications \cite{rontani}

The experimental realization of a bosonic ground state has concentrated  on excitons in semiconductors. However, in this class of materials the
excitons are not in the energetically lowest state but laser excited particles. A steady state equilibrium between optical generation of the
excitons and the decay within their life time  must be achieved \cite{littlewood}. Our approach has been the application of pressure in order to
lower and fine-tune the energy of the excitons to the energy level of electrons in the valence band (see Figure 1).

TmSe$_{0.45}$Te$_{0.55}$ is a narrow gap semiconductor. Chemistry would suggest the valences Tm$^{+2}$[Se$_{0.45}$Te$_{0.55}$]$^{-2}$ for the
ionic binding regime but, actually, the small energetic separation of the 4f states from the conduction band causes a 4f-5d hybridization with
the result of a so-called intermediate valent state Tm$^{+2.2}$ where the generally localized 4f states acquire a narrow (some tens of meV) wide
4f valence band. At 300 K the energy gap closes on applying pressure \cite{neuenschwander} and there is a transition from a semiconductor to a
ferromagnetic semimetal. But at temperatures below 250 K, the resistivity increases strongly in the pressure range between 5 and 8 kbar, despite
of the anticipated closing of the gap. Hall effect measurements \cite{bucher1} reveal that the anomaly is due to localization of electrons
(inset to Fig. 2\,a) as demanded by a formation of excitons with their energy level crossing the high density of states of the quasi-localized
4f electrons around 5 kbar (Fig. 1).

\begin{figure}[h] \label{fig1}
    \includegraphics[width=80mm]{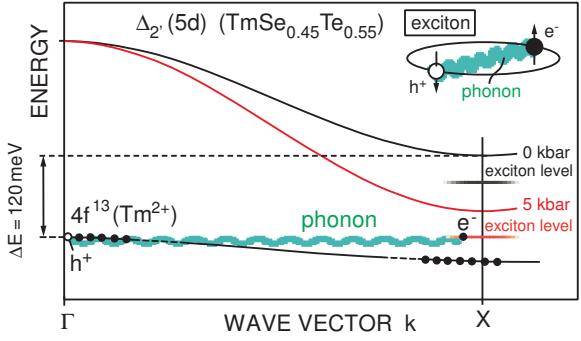}
\caption{Schematic illustration of the band structure of TmSe$_{0.45}$Te$_{0.55}$ on applying external pressure. Exciton levels below the $X$
point of the Brillouin zone reduce  their separation to the quasi-localized 4f$^{13}$ electrons of Tm$^{2+}$. Around 5 kbar, it becomes
energetically favorable for an electron to occupy the excitonic energy level, leaving behind a positively charged hole. Conservation of momentum
requires the binding of a phonon with a wave vector from the $\Gamma$ to the $X$ point. }
\end{figure}

Further hints to a new phase have come from heat transport. Thermal conductivity \cite{bucher2} showed a first order jump, like the resistivity,
giving heat conductivity below 20 K that is higher than in the semimetal state. A phase diagram has been deduced out of these data of the
transport properties (Fig. 2a). Yet the formation of a new phase must be corroborated by thermodynamics. This was the motivation to measure the
heat capacity and to establish the existence of a new thermodynamical phase.


A self-clamping CuBe pressure cell has been   used with a silicon-based pressure transmitting fluid. A superconducting lead manometer was
employed as the pressure gauge. The pressure was not constant on cooling down but rather follows a pressure path that depends on the cell  and
the starting pressure.  \cite{thompson}. An ac calorimetry technique \cite{seidel} was adapted  for studying phase transitions at hydrostatic
pressure. With this technique,  a small sinusoidal heat input is supplied to the sample (1x1x0.2 mm$^3$) through a constantan  wire bent to a
meander-like structure attached to one side of the sample. The resulting temperature oscillation was detected by a Chromel-Au/Fe thermoelement
attached to the other side of the single crystal. At steady state, the time-average temperature $T_1$ of the sample is larger than the bath
temperature (pressure medium). In addition, there is a small temperature oscillation about $T_1$ with a peak-to-peak temperature  $\Delta
T_{ac}$. Under optimal conditions of the ac frequency, $\Delta T_{ac}$ is directly related to the heat capacity. However the calculation of the
absolute value of the heat capacity would require the knowledge of thermal resistances to the bath and the thermometer. In the absence of this
information, the absolute value of heat capacity was estimated by scaling  the high-temperature ac heat capacity to the Dulong-Petit value of
TmSe$_{0.45}$Te$_{0.55}$ of 52 J/mol\,K per f.u. However, some influence of the pressure transmitting fluid can not be prevented; thus,
conclusions are based mainly on differential changes of the heat capacity. With this technique, phase transitions also manifest themselves
through a marked phase shift of the thermo-signal across the sample.


\begin{figure}[th] \label{fig2}
\includegraphics[width=80mm]{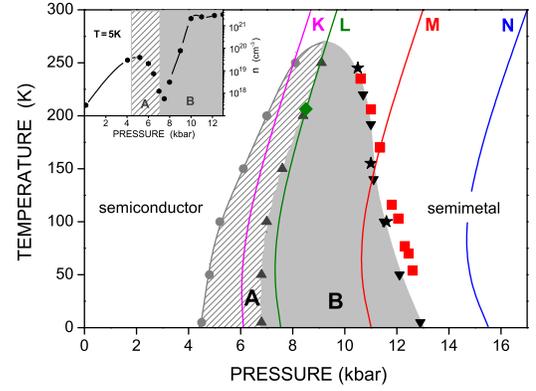}
\includegraphics[width=80mm]{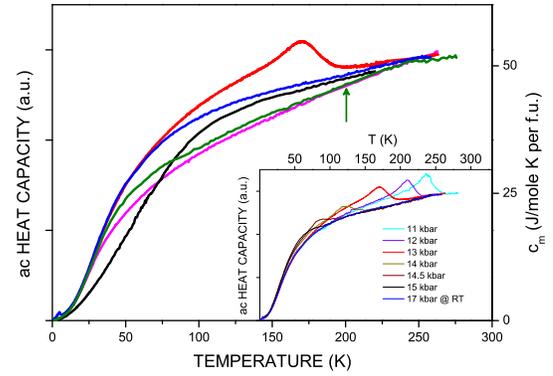}
\caption{Heat capacity of TmSe$_{0.45}$Te$_{0.55}$ at high  pressures  a) Phase diagram constructed from a number of different measurements:
{\textcolor[rgb]{0.50,0.50,0.50}{$\bullet$}} (gray) onset of resistivity increase; $\blacktriangle$ maximum of resistivity;
{\textcolor[rgb]{1.00,0.00,0.00}{$\blacksquare$}} maximum of ac heat capacity; $\bigstar$ first order transition of heat  conductivity;
$\blacktriangledown$ first order transition of resistivity and {\color{green}$\blacklozenge$} the onset of deviation of ac heat capacity of
pressure run L. The inset shows the pressure-dependent Hall number $n$ at 5 K \cite{bucher1}, where $n$ decreases in phase A and electrons start
to delocalize again in phase B. b) ac heat capacity data for different pressure runs: black (zero pressure), red (run M), blue (run N,
metallic), green (run L) and magenta (run K). A ferromagnetic phase transition at 6 K is discernible in the metallic phase. The inset to Fig. 2
b) represents the experimental results for seven pressure runs that cross the semimetal to phase B boundary. Peaks in $C_{ac}$ ({\color{red}$
\blacksquare$}) are denoted in Fig. 2 a). The green arrow indicates the phase transition from phase A to phase B. }
\end{figure}


Figure 2 b) shows the ac heat capacity $C_{ac}$ for selected pressure runs. The pressure inside the self-clamped cell reduces on cooling, as
determined in independent measurements, and, consequently, the heat capacity could be measured neither at constant pressure nor at constant
volume. Depicted are the ac heat capacities of representative pressure runs K, L, M, N and zero pressure (black). The heat capacities show a
conventional curvature for a Debye-like phonon density of state at zero pressure and in the metallic regime. The phase boundary from
semiconducting to phase A is reflected as a large change in slope of the heat capacity of run K around 250 K, signalling a second order phase
transition. The reduced heat capacity of run K below 250 K (20 meV) indicates thermally no longer activated lattice vibrations. The
quasi-localization of free carriers starts below the same temperature \cite{neuenschwander}. Indeed, wave-vector conservation upon the formation
of excitons in an indirect gap semiconductor implies the binding of high energy phonons at the boundary of the Brillouin zone to form excitons
(Fig. 1). The number of excitons is then given approximately by the decrease in free-carrier density, measured by the Hall number
\cite{bucher1}: $\Delta n_{max} \simeq 10^{20}$ e$^-$/cm$^3$. The heat capacity of run K is nearly linear between 100 to 250 K, which points to
a constant phonon density of states. The reduced heat capacity also implies a stiffer lattice \cite{barron}, \cite{bucher2}.


Pressure run L enters phase B from phase A around 200 K. Only 1 kbar higher in the semimetal phase, the heat capacity changes its shape
qualitatively (inset to Fig. 2b). Pressure run M enters phase B from  the semimetal side and a giant change takes place below 200 K. In the
metallic region (run N) the heat capacity recovers a conventional Debye-like curvature but with a lower Debye temperature $\Theta_D$, as
expected for intermediate valent systems at higher pressures \cite{bucher2}.

By subtracting the heat capacity of run N, the specific heat of the runs from the  metallic region into phase B can be decomposed into a
phononic part and a contribution $\Delta C$ belonging to the phase transition. A fit of the extra heat capacity $\Delta C$ reveals a Gaussian
peak together with a broader contribution at lower temperatures (Fig. 3). A Gaussian shape of the heat capacity is known for first order
transitions \cite{schilling} as a temperature gradient is always present \cite{firstorder}. Furthermore, neutron measurements
\cite{neuenschwander} show an isostructural phase transition with an expansion of the volume by 4.8\% at the phase boundary from semimetal to
phase B (inset to Fig.3). This expansion of the lattice increases the pressure inside the self-clamped pressure cell, counterbalancing the phase
transition. A first order transition also has been observed in the resistivity \cite{neuenschwander}, Hall constant  \cite{bucher1} and
 heat conductivity \cite{bucher2}.

\begin{figure} \label{fig3}
  \includegraphics[width=80mm]{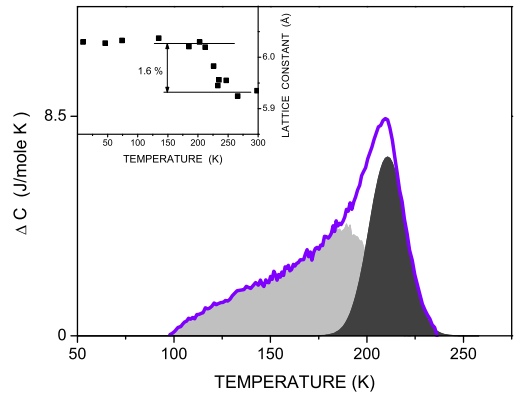}
\caption{Heat capacity  for the violet pressure run (11.8 kbar at 5 K, Fig. 2b) after subtracting the conventional background characterized in
run N. A fit of the extra heat capacity $\Delta C$ reveals a Gaussian peak (dark gray) together with a broader contribution at lower
temperatures (light gray). The dark gray area reflects a latent heat.  The accompanying expansion of the lattice increases the pressure inside
the self-clamped pressure cell counterbalancing the phase transition. The inset shows the large lattice expansion at the same pressure
\cite{neuenschwander}.}
\end{figure}

The broadening of the transition allows calculation of the  entropy change and, therefore, a verification of the
 Clausius- Clapeyron relation:
\begin{equation}\label{clausius}
    \frac{dp}{dT} = \frac{\Delta S}{\Delta V} = \frac{\Delta Q}{T \cdot \Delta V}
\end{equation}

The slope $\frac{dp}{dT}$ on the left side of Eq. (\ref{clausius}) can be determined from the phase
boundary shown in Fig. 2a and gives $ - 5.2\cdot 10^5  \hbox{ J/m$^3$ K}$ at 200 K. For the right
side of Eq. (1), the corresponding change of entropy at the boundary (dark gray shaded peak in Fig.
3) provides $\Delta S$=0.8 J/mol K. With a volume expansion of $1.56\cdot10^{-6}\,$m$^3$/mol, the
resulting $\Delta S/\Delta V = - 5.1\cdot 10^5 \hbox{ J/m$^3$ K}$ is in excellent agrement with the
slope of the phase boundary. The  additional broad heat capacity at lower temperatures comprises
$\Delta S$=1.6 J/mol K and, hence, gives a total change of entropy of 2.4 J/mol K (inset to
Fig.\,4).

\begin{figure} \label{fig4}
  \includegraphics[width=80mm]{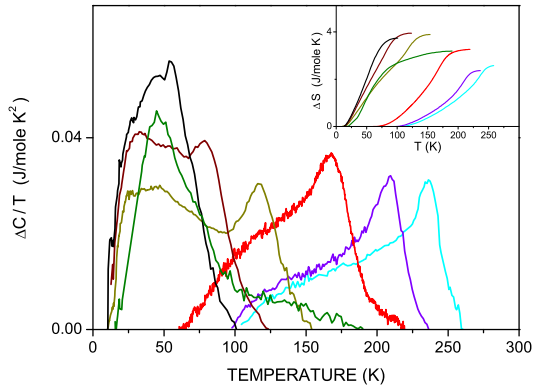}
\caption{ $\Delta C /T$ for different pressure runs.  The pressures at low temperatures are from right to left: cyan (10.7 kbar), violet (11.8
kbar), red (11.9 kbar), dark yellow ( 13.5 kbar), brown (13.7 kbar), black (13.9 kbar) and green (7.6 kbar). The change  of entropy due to
entering phase B from phase A was calculated by subtracting the heat capacity of run K from run L (green curve). The inset shows the integrated
area of $\Delta C /T$, which corresponds to the involved change of entropy and is about the same for all pressure runs. }
\end{figure}

Figure 4 shows the additional  contribution $\Delta C/T$ for the pressure runs into phase B. The
change of entropy due to entering phase B from phase A was estimated by subtracting the heat
capacity of run K from run L (green curve in Figure 4). The total change of entropy is similar
for all runs.


Let us first discuss the first order phase transition into phase B. The spin entropies for the
thulium ions are $S_{spin}/k_B = \ln (2J+1) = 2.08$ for Tm$^{+2}$ ($J=\frac{7}{2}$) and 2.56
for Tm$^{+3}$ ($J= 6$), respectively. A gain of entropy by a valence change from Tm$^{+3}$ to
Tm$^{+2}$ would result in $\Delta  S_{spin} = 0.48 \cdot $k$_B$ = 3.98 J/mol K , in good agreement
with the experimental result shown in the inset of Fig. 4. The first order expansion of the lattice might then solely be a
consequence of the larger Tm$^{+2}$ ions.
The f-electron intra-site Coulomb repulsion does not favor this picture and, particularly, the Hall
measurements show that electrons in phase B are not localized as they are in phase A \cite{bucher1}.
The electrons in phase B seem to compensate the local magnetic moment of the Tm ions but are not
localized on the Tm site itself. The idea of electrons redistributed over d-orbitals that are bound
to its original anion site, i.e. excitons, also has been used to explain the golden phase of SmS
\cite{sato},\cite{kikoin}.

The expansion of the crystal at the first order phase transition against the pressure medium
($\approx$ 12 kbar) requires an energy $\Delta E_{elastic} = p \cdot \Delta V \approx 1800$ J/mol =
20 meV/f.u. This value is to be compared with the latent heat $L = \Delta S \cdot T = $ 165 J/mol =
1.7 meV/f.u. The spin entropy which usually drives the Kondo and Kondo - lattice effects in highly
correlated electron systems can not explain the phase transition. Only an electronic energy could
account for the large energy per formula unit; the binding energy of excitons is in this range.
This formation of excitons has been predicted for electrons in TmSe$_{0.45}$Te$_{0.55}$ interacting
via a statically screened Coulomb potential \cite{bronold1}. The broad contribution to the heat
capacity in phase B (Fig. 3) could be attributed to the continuous formation of bound excitons. But
the formation of a large number of excitons with heavy holes is subject to crystallization
\cite{rice} and the first order transition could indicate a melting process.

The localization of the electrons in phase A \cite{bucher1} is accompanied by the disappearance of
phonons (below 250 K of run K in Fig. 2b) as expected at the formation of a large number of
excitons in an indirect gap semiconductor. Hence, the insulating phase A  could be interpreted by
the opening of a "many-body gap", i.e. the proposed Bose condensate of excitons \cite{sham1}. The
ground state of an excitonic insulator in an indirect gap semiconductor competes with an
instability to an electron - hole liquid \cite{rice}. But the asymmetry of electron to hole mass of
$m_h/m_e \approx 80$   may prevent the e-h liquid instability in TmSe$_{0.45}$Te$_{0.55}$ and
favors crystallization in a two-component Coulomb system \cite{fehske1}. Even more significant
might be the fact that the movement of holes is inherently related to a change of the local
magnetic moment.  A magnetic binding energy may then support the formation of hole - electron
bosons.
\\

In summary ac heat capacity measurements of TmSe$_{0.45}$Te$_{0.55}$ at high pressure have revealed a pronounced change in the phonon spectrum
up to 250 K. The binding of a large number of high energy phonons is in agreement with the formation of an excitonic insulator phase. The gain
of spin entropy points to the formation of singlet excitons.

\begin{acknowledgments}
 The authors thank F. Bronold, H.Fehske, and A. Schilling for valuable discussions.
 Work at Los Alamos was performed under the auspices of the US DOE, Office of Science.
 Correspondence and requests for further information.
 should be addressed to Benno Bucher~(email: bbucher@hsr.ch).
\end{acknowledgments}



\bibliography{prlreferenzen}

\begin{references}{ }
\bibitem{firstorder} This phenomenon is similar to boiling a pot of water: the water does not instantly turn into gas, but
forms a turbulent mixture of water and water vapor bubbles. Furthermore if a substance as water and TmSe$_{0.45}$Te$_{0.55}$ expands upon
cooling the effect called regelation takes place: melting under pressure and freezing again when the pressure is reduced.
\end{references}

\end{document}